\documentclass[aps,prb,twocolumn,floatfix] {revtex4-2}
\usepackage{graphicx,amsfonts}
\usepackage{bm,color}
\usepackage{multirow}
\usepackage{amssymb,amsmath,hyperref}
\usepackage{wrapfig}
\usepackage{lineno}
\usepackage{float}
\usepackage{xcolor}
\usepackage{comment}

\DeclareUnicodeCharacter{2212}{-}
\usepackage{dcolumn}
\usepackage{bm}
 
\newcommand{\cred}[1]{{\color{red} #1}}

\begin{document}

\title{Strong-coupling quantum thermodynamics using a superconducting flux qubit}

\author{Rishabh Upadhyay  $^{1,2}$ }\email{rishabh.upadhyay@aalto.fi}

\author{Bayan Karimi$^1$}

\author{Diego Subero$^{1}$}

\author{Christoforus Dimas Satrya$^{1}$}

\author{Joonas T. Peltonen$^1$}

\author{Yu-Cheng Chang$^1$}

\author{Jukka P. Pekola$^1$}

\affiliation{$^1$ Pico group, QTF Centre of Excellence, Department of Applied Physics, Aalto University School of Science, P.O. Box 13500, 00076 Aalto, Finland}

\affiliation{$^2$ VTT Technical Research Centre of Finland Ltd, Tietotie 3, 02150 Espoo, Finland}

\begin{abstract}

Thermodynamics in quantum circuits aims to find improved functionalities of thermal machines, highlight fundamental phenomena peculiar to quantum nature in thermodynamics, and point out limitations in quantum information processing due to coupling of the system to its environment. An important aspect to achieve some of these goals is the regime of strong coupling that has remained until now a domain of theoretical works only. \textcolor{black} {Our aim is to demonstrate strong coupling features in heat transport using a superconducting flux qubit, which is capable of reaching strong to deep-ultra strong coupling regimes, as shown in previous studies. Here, we show experimental evidence of strong coupling by observing a hybridized state of the qubit with two cavities coupled to it, leading to a triplet-like thermal transport via this combined system around the minimum energy of the qubit, at power levels of tens of femtowatts, exceeding by an order of magnitude those in earlier experiments. We also demonstrate close to 100\% on-off switching ratio of heat current mediated by photons by applying magnetic flux to the qubit.} Our experiment opens a way towards testing debated questions in strong coupling thermodynamics such as what heat in this regime is. We also present a theoretical model that aligns with our experimental findings and explains the mechanism behind heat transport in our device. \textcolor{black} {Furthermore, our experiment opens new possibilities for quantum thermodynamics, aiming to realize true quantum heat engines and refrigerators with enhanced power and efficiency, by leveraging ultra-strong coupling between the system and its environment in future experiments.}

\end{abstract}

\maketitle{}

\section{Introduction}
\vspace{5mm}

\textcolor{black}{Quantum thermodynamics (QTD) \cite{Pekola,Deffner,Pekola JP,Sai}, or thermodynamics of quantum systems and processes, has strong roots in the theory of open quantum systems. Experimental work on QTD is limited, since access to true thermodynamic quantities, most importantly heat, is challenging due to the smallness of energies involved in these sensitive quantum systems~\cite{Francesco Giazotto,Ronzani2019,F.Giazotto,Mikko-QCR,D.Subero,Jorden}. Therefore most experimental works are only indirect, in that they address observables not directly related to thermodynamics, like populations in the quantum systems themselves, and infer thermodynamic quantities like heat, work and efficiencies based on the models believed to govern the system and its environment. Superconducting quantum circuits, combined with electronic on-chip reservoirs form an exception in this respect, since heat currents can be measured directly at low temperatures via thermometry, due to the efficient isolation and small heat capacities of the reservoirs~\cite{F.Giazotto,Mikko-QCR,D.Subero}.} Moreover, thanks to the rapid development of superconducting quantum circuits over the past few decades ~\cite{Kjaergaard,JKoch,Wallraff,JohnClarke,Morten}, they can be considered as next to ideal coherent quantum systems that then couple to the heat reservoirs; furthermore, accurate design and control of the circuits is achievable. 

\begin{figure*}[ht]
\centering
\includegraphics [width = 0.65\textwidth] {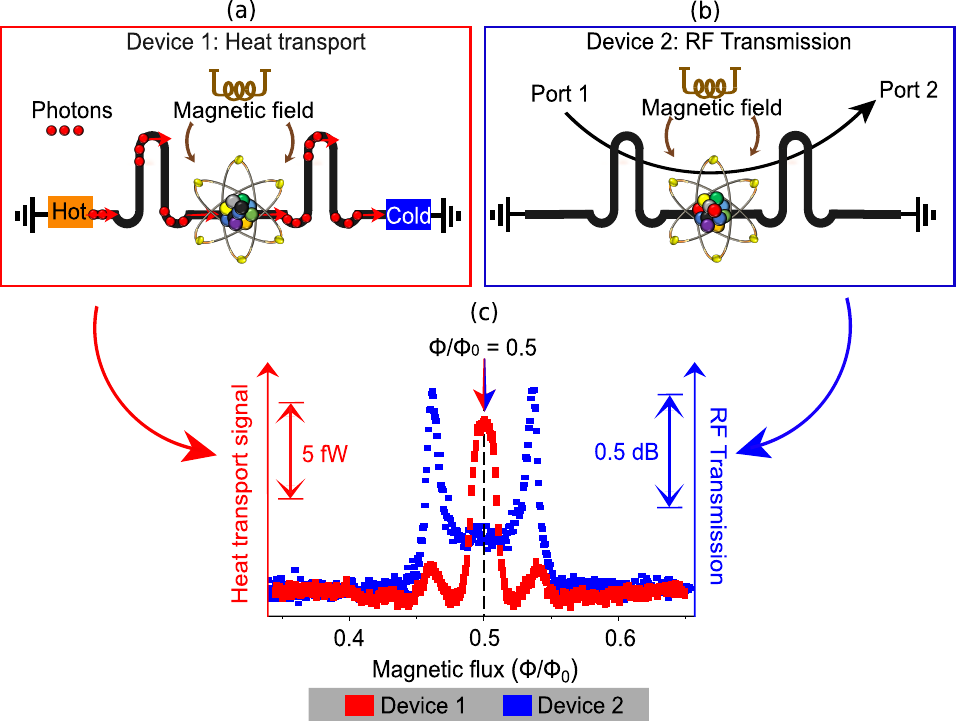}
\caption{
{\bf Device concept.} 
\textcolor{black}{(a) A schematic showing thermal photons (red) generated by Joule heating in a metallic reservoir being blocked by an artificial atom as they travel toward the cold reservoir. The atom, tuned by a magnetic field, allows photon passage only close to half-flux quantum (indicated by red arrows). (b) Scheme for a two-port readout of a transmission signal in a spectroscopy enabled device without any metallic reservoirs. (c) A qualitative, comparative plot shows maximum heat transport signal (red) at half-flux quantum, with suppression when qubit and resonator frequencies align, as seen in the transmission signal (blue). The power scales on both x-axes are provided for approximate reference and are not intended for direct comparison.} Here, $\Phi$ is the applied flux, and $\Phi_0 = h /2e$ is the flux quantum.}
\label{working concept}
\end{figure*}

On the way of realizing true quantum heat engines and refrigerators~\cite{B.Karimi,Kosloff,Erdman,Menczel}, one needs to understand and optimize heat transport mediated by a quantum circuit, which is to be the working substance of the quantum thermal machine. Here the important aspect is the coupling between the quantum circuit, in most cases a qubit, and the reservoir ~\cite{B.Karimi, Ronzani2019}. \textcolor{black}{The whole process of heat transport via the qubit is based on sequential excitation $g \rightarrow e$ and relaxation $e \rightarrow g$ events, where $e$ and $g$ represents exited and ground states of the qubit.} Excitation removes heat equal to the qubit energy separation from the hot bath, and relaxation releases the same amount of heat to the cold bath ~\cite{B.Karimi, Ronzani2019}. Superconducting transmon ~\cite{JKoch,JohnClarke,Morten,Majer} is the most studied qubit, but like in quantum information processing, also in thermal transport it has its limitations. To retain the anharmonicity of the transmon, an important character of the qubit, it can be coupled only weakly to the remaining circuit ~\cite{Morten,Majer}. On the other hand, the flux qubit~\cite{Orlando, Mooij, Yan.F}, particularly in a galvanically coupled configuration~\cite{J.Bourassa, T.Miyanaga, Niemczyk, R.Upadhyay}, enables access to the strong and ultra-strong coupling regimes, as well as the deep ultra-strong coupling regime~\cite{Yoshihara.F, F.Yoshihara}. \textcolor{black}{These coupling regimes open the way of increased powers in thermal machines, and on the fundamental side gives the possibility to address what is called strong coupling thermodynamics ~\cite{Falci G, Giannelli L, B.Karimi, Dou, Perarnau, Rivas, Talkner, Teufel, Yoshihara, Choi, Huang, Ding, Bosman}, but has been out of reach from experiments up to now.} From the conceptual point of view, basic questions such as what heat is in the case when system and reservoir are strongly coupled, is an important issue: in the weak coupling theories the energy associated to the coupling Hamiltonian is ignored which is not any more possible for a strongly coupled system, such as the reported one.

\begin{figure*}[ht]
\centering
\includegraphics [width = 1\textwidth] {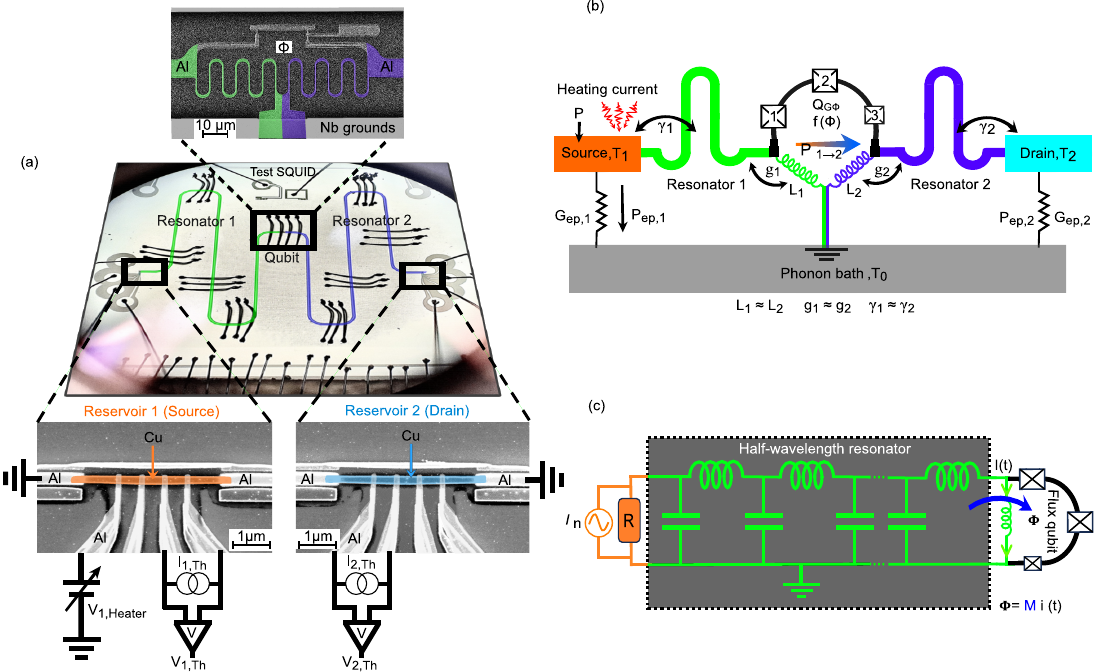}
\caption{{\bf Device design and circuit.}
(a) An optical image of the reported device. At the center, within the solid black box, is a superconducting qubit coupled to Resonator 1 (green) and Resonator 2 (blue) and connected to the ground plane. The lower sections display Reservoir 1 and Reservoir 2, along with a brief measurement scheme provided in detail in the article Appendix B. \textcolor{black}{(b) Conceptual schematic showing both reservoirs, each connected to a phonon bath at temperature $T_{0}$, with thermal conductance $G_{ep,1}$ and $G_{ep,2}$, respectively. The qubit's flux-dependent frequency $(f(\Phi))$ controls heat transport ($Q_{G\Phi}$), thus it acts as a tunable quantum heat switch.} (c) Circuit diagram shows a resistor $(R)$ connected to a half-wavelength resonator and qubit, with $i_n$ as current noise and $\Phi = Mi(t)$ for input coil inductance. Reported SEM images have been colored for clarity.}

\label{e-pg_coupling}
\end{figure*}

\textcolor{black}{In this work, we report on two devices: Device~1 and Device~2. Device~1 is a thermal device architecture that forms the core of the reported experiment. Device~2, which plays a supporting role, is a spectroscopy-enabled device designed for performing qubit spectroscopy useful for complete understanding of the experiment. In Device 1, we experimentally investigate quantum heat transport between two reservoirs mediated by an interconnected artificial atom, which is a superconducting flux qubit.} The concept of Device 1, illustrated in Fig. \ref{working concept} (a), depicts photons energized by Joule heating in a metallic reservoir transferring heat to a cooler region, but being blocked by a superconducting qubit, which, modulated by a weak magnetic field, acts as a controlled gate, allowing photon passage only at certain flux values. \textcolor{black}{By varying the external magnetic field, we tune the qubit frequency accordingly with respect to that of the cavity, which allows us to control heat transport across the qubit \cite{Ronzani2019, Subero}.} The observed heat transport exhibits a triplet structure, with the central peak reaching its maximum at the degeneracy point, corresponding to half a flux quantum in a superconducting flux qubit, as depicted in Fig. \ref{working concept} (c). \textcolor{black}{In an RF environment, utilizing a read-out resonator, characteristic transmission across two ports of a spectroscopy enabled device shown in panel (b) of Fig. \ref{working concept}, is observed upon changing the magnetic field, as observed in Fig.~\ref{working concept}\,(c).} By overlaying these two signals, a preliminary understanding of the system's behavior under varying magnetic field in both DC and RF environments is achieved, revealing the origin of the satellite peaks in the triplet heat transport signal when the qubit frequency aligns with the resonator frequency, outlined in Fig.~\ref{working concept}\,(c). In Ref.  \cite{Ronzani2019}, a weakly-anharmonic transmon qubit was used to demonstrate the concept of such a measurement, however the device performance was limited by its weak qubit-resonator ($g$) and resonator-reservoir ($\gamma$) coupling strengths. \textcolor{black}{The reported flux qubit architecture enables access to ultra-strong and even deep-ultra-strong coupling regimes with minimal modifications to the qubit design. These regimes, which remain experimentally unexplored in the field of quantum thermodynamics, are largely inaccessible using transmon-type qubit architectures. The novelty of our approach lies in exploiting the length-dependent mutual inductance of the coupling element, which enables further enhancement of the qubit–resonator coupling strength, pushing it into the ultra-strong coupling regime and beyond \cite{R.Upadhyay, J.Bourassa, Niemczyk, Yoshihara.F}.} A theoretical perspective towards high performance of such devices is addressed in Ref. \cite{B.Karimi}, where the importance of couplings  $g$ and $\gamma$ is discussed. \textcolor{black}{In this work, within the framework of strong coupling between the reservoir, resonator and qubit, our results demonstrate an order of magnitude increase in power modulation along with a nearly 100\% on-off switching ratio of heat current in the photon dominated heat transport regime.} With a primary focus on experimental observations, this work also introduces a theoretical model that quantitatively captures the observed heat current and qualitatively explains the flux-dependent power. The device design is reported in Fig. \ref{e-pg_coupling}, with detailed description provided in Section II of this article.

\vspace{2mm}

\section{Device design}
\vspace{2mm}

\label{Setup}
\vspace{2mm}

\textcolor{black}{Since the primary experiment is performed using Device~1, we focus on its design in this section. The design details of Device~2 are discussed separately in Section~IV.} \textcolor{black}{The hybrid quantum system is composed of a flux qubit galvanically connected to two nominally identical superconducting half-wavelength coplanar resonators with designed frequency of $f_{r} \approx 6.5$ GHz, which are in turn coupled to two copper reservoirs referred to as Source (S) and Drain (D) reservoir.} Thus, the device architecture can be succinctly described as: Source -- Resonator 1 -- Flux qubit -- Resonator 2 – Drain. For convenience, we will refer to the Source reservoir as Reservoir 1 and the Drain reservoir as Reservoir 2 throughout the article. \textcolor{black}{The measured resistance of the reservoirs $R$ are $\approx 6~\Omega$ at helium temperatures.} \textcolor{black}{The reservoirs are coupled to their respective resonators with a coupling strength given by $\gamma/2\pi = f_{r}/Q_{R}$ $\approx 500$~MHz. Here, $Q_{R}$ is the quality factor of the reservoir given by  $Q_{R}= \frac{\pi}{2}Z_{0}/R$, where $Z_{0} (\approx50 \Omega)$ is the designed characteristic impedance of the resonator~\cite{Ronzani2019}.} The measured device is shown in Fig.~ \ref{e-pg_coupling}\,(a). The flux qubit in the middle is a superconducting loop interrupted by three Josephson junctions in series where two of them are designed to be identical with the critical current $I_{C2}=I_{C3}=I_{C}$, while the area of the first junction is reduced by a factor of $\alpha\approx 0.6$ with critical current $I_{C1}=\alpha I_{C}$. \textcolor{black}{The scanning electron microscope (SEM) image} of the zoom-in of the flux qubit in Fig.~ \ref{e-pg_coupling}\,(a) shows the galvanic connections to Resonator 1 and Resonator 2 via the inductances $L_1$ and $L_2$ realized by a thin superconducting Al wire, facilitating couplings $g_1$ and $g_2$ with each resonator, respectively. The coupling constants are proportional to the corresponding inductances as $g_j\propto L_j$ (where $L_1 \approx L_2$)~\cite{J.Bourassa, T.Miyanaga}. The fundamental frequencies of the resonators and the qubit-resonator coupling have been carefully designed to be identical on both sides of the qubit (side 1 and side 2) to maintain symmetry in the device. The estimated coupling strength $g_1/2\pi\simeq g_2/2\pi\simeq 200$ MHz, was extracted by fitting spectroscopy data described in section IV.

Each of the $\lambda/2$ resonators attached to the flux qubit is terminated at its end by a reservoir. They are shown with orange and black in the bottom of Fig. \ref{e-pg_coupling}\,(a). Each reservoir has four normal metal -- insulator -- superconductor (NIS) junctions attached to it. The electronic temperatures of the Reservoir 1, $T_{\rm 1}$, and Reservoir 2, $T_{\rm 2}$ can be controlled (by voltage biasing NIS as heater) and monitored (via SINIS junctions acting as thermometers) as shown in the SEM images of Fig. \ref{e-pg_coupling}\,(a), respectively ~\cite{Francesco Giazotto}. Each reservoir is strongly coupled to its respective resonator. The thermal model of the device including both reservoirs, connected via a tuneable hybrid quantum system (Resonator 1– Flux qubit – Resonator 2) and each of them furthermore connected to the phonon bath at temperature $T_{0}$, is schematically shown in Fig. \ref{e-pg_coupling}\,(b). \textcolor{black}{Panel (c) of Fig.~\ref{e-pg_coupling} shows the circuit diagram of a single reservoir coupled to the resonator-qubit system, which is used to obtain the noise spectrum $S_\Phi^{(r)}(\omega_{ij})$ for the theoretical analysis.}


\begin{figure*}[ht]
\centering
\includegraphics [width = 0.7\textwidth] {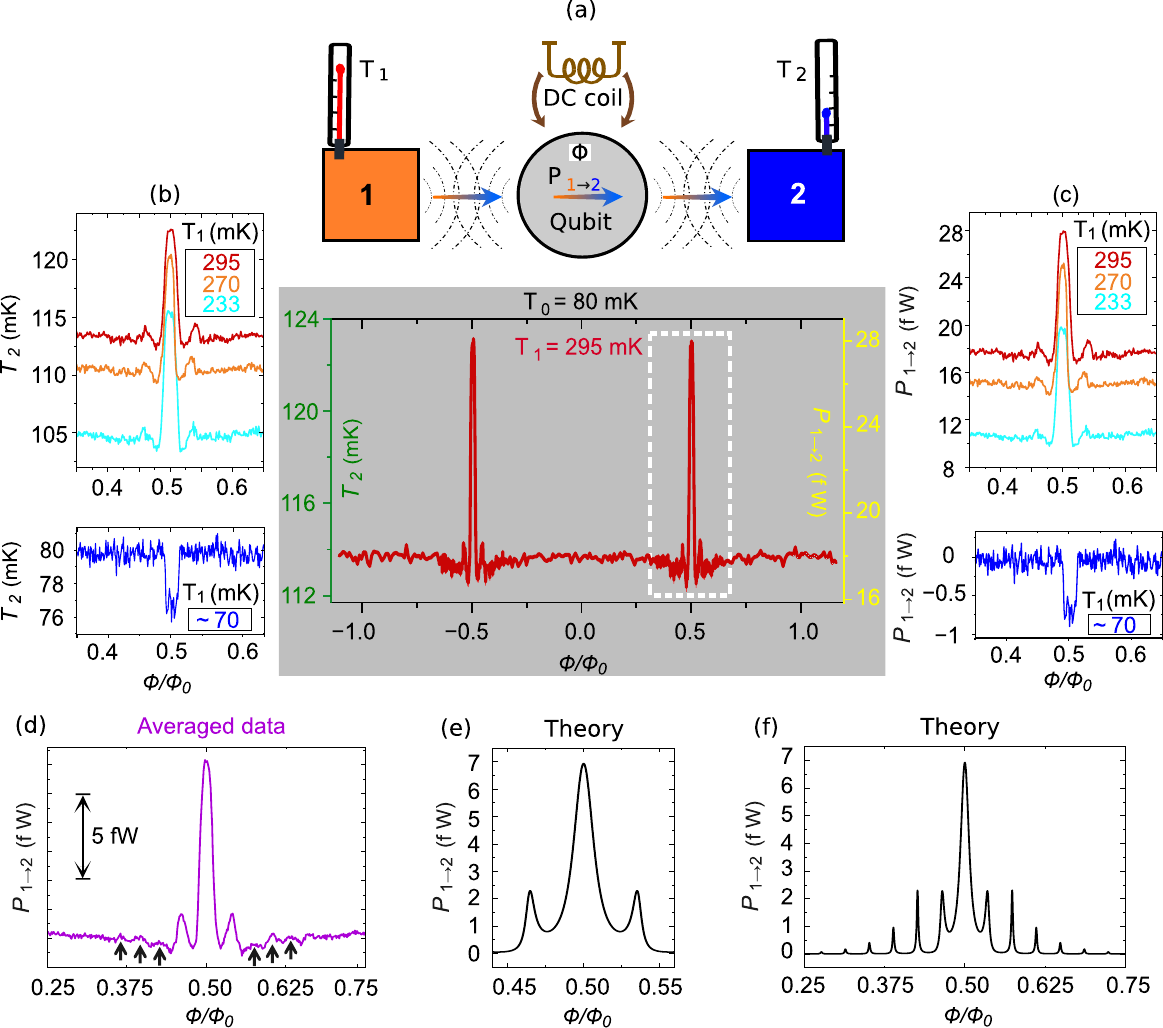}
\caption{ {\bf Flux-dependent power transport.} (a) Flux-dependent power transport, $P_{1 \rightarrow 2}$, across the qubit, driven by a magnetic flux generated by the coil, with power from Reservoir 1 (red) absorbed by Reservoir 2 (blue), at $T_{0}= 80\,$mK. The bottom plot shows the modulation of Reservoir 2's temperature (left y-axis) and transported power (right y-axis) as a function of the applied magnetic field. \textcolor{black}{The region within the dashed rectangle in the center of the image is enlarged in panels (b) and (c). In panel (b), the y-axis represents temperature, while in panel (c), it represents the power absorbed by Reservoir 2. (b) $T_{2}$ vs applied magnetic flux at different $T_1$.} (c) $P_{1 \to 2}$ vs applied magnetic flux at different $T_1$. (d) Averaged data (pink curve) from multiple sweeps shows small satellite peaks (black arrows). (e) $P_{1 \to 2}$ from the theoretical model. (f) Satellite peaks captured by the theoretical model. \textcolor{black}{The average RMS noise, calculated over the range $\pm 0.25 \, \Phi/\Phi_0$ for the data shown in panel~(a), is $\approx~$0.14~fW, which is considerably smaller than the amplitude of the signal of interest.}}

\label{dc_results}
\end{figure*}

\begin{figure}[ht]
\centering
\includegraphics [width=0.70\columnwidth] {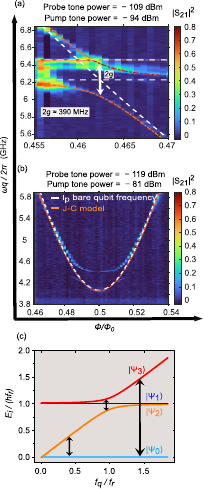}
\caption{{\bf Device spectroscopy.} \textcolor{black}{(a) Two-tone spectroscopy measurement probing the interaction of a strongly coupled qubit--resonator system via a read-out resonator at the readout frequency \( f_{\text{read-out}} \approx 7\,\text{GHz} \). The estimated coupling strength is \( g/2\pi \approx 200\,\text{MHz} \), obtained by fitting with the J-C model. (b) Qubit energy spectrum vs applied magnetic flux, with minimum frequency ($\approx 4.0$ GHz) at half-flux. The orange dashed-line shows the J-C model fit; the white dashed-line fits $\hbar \, \epsilon =  2I_{p}\,(\Phi-\Phi_{0}/2)$, with $I_p \approx 21$ nA. The avoided crossing with a stationary feature around $\approx 4.4~\mathrm{GHz}$ arises from the interaction of the system with a parasitic resonance which is disregarded in the analysis of the results. (c) Dominant transitions from the theoretical model responsible for heat transport across the qubit.}}

\label{spec_results}
\end{figure}

\vspace{2mm}
\section{Thermal transport measurements}

\textcolor{black}{The heat is transported from the Reservoir 1 to Reservoir 2 under steady-state conditions. Energy conservation for the Reservoir 2 reads $P_{1\rightarrow 2}= P_{\rm ep,2 }$, where $P_{\rm ep,2}= \Sigma \mathcal{V}\,(T_{\rm 2}^{n} - T_{\rm 0}^{n})$ is the conventional power between the electrons in the Reservoir 2 at temperature $T_2$ and the phonon bath at temperature $T_0$, see Fig.~\ref{e-pg_coupling} (b). Here, indices 1 and 2 denote the respective reservoirs, and `ep' refers to electron-phonon coupling ~\cite{Pekola JP}.} Here, $\Sigma$ represents the electron-phonon coupling constant, $\mathcal{V}$ the volume of the reservoir, and the phonon temperature $T_{\rm 0}$ is equal to the cryostat temperature. \textcolor{black}{The constant value of \( n \), is material-specific and depends on the type of disorder in normal metals. For a clean, three-dimensional normal metal, \( n = 5 \) \cite{Sergeev, Wellstood, Wang}.} The equation can be linearized for small temperature differences to $P_{\rm ep,2} = G_{\rm ep,2 }\Delta T$ with $ G_{\rm ep,2}= n \Sigma \mathcal{V}T_{\rm 2}^n$ denoting the electron-phonon thermal conductance. \textcolor{black}{For the reported experiment, $ \Sigma \mathcal{V} $ for Reservoir 1 and Reservoir 2 are estimated to be $ 9.35 \times 10^{-10} $ and $ 11.44 \times 10^{-10} $ WK$^{-5}$, respectively. This can be calculated using the relation $ \frac{1}{2}I^{2}R_{\rm Heater} = \Sigma \mathcal{V} (T_{e}^{n} - T_{0}^{n}) $, where $I$ is the applied current, $R_{\rm Heater}$ is the resistance of heater junction which was measured $\approx 20\, k\Omega$, $T_{e}$ is the electronic temperature, and the slope corresponds to $ \Sigma \mathcal{V} $~\cite{Wang}.}

We monitor $T_{\rm 2}$ when heating the Reservoir 1 by Joule power applied via a single NIS junction  ~\cite{Francesco Giazotto}. The temperature, $T_{\rm 2}$, is monitored via measurement of the thermal voltage across another pair of NIS junctions attached to it and  $T_{\rm 1}$ is monitored similarly. The calibration process for converting the thermometer's voltage into temperature, and subsequently into absorbed power, is detailed in Appendix B. Another control parameter in the experiment is the magnetic flux on the qubit affecting the energy separation between the ground and excited states. Schematics of the experiment are shown in Fig.~\ref{dc_results}\,(a). 

While varying the magnetic flux with constant temperature bias between Reservoir 1 and Reservoir 2, the measured  $T_{\rm 2}$ exhibits variations periodic in superconducting flux quantum $\Phi_0=h/2e$, as plotted in Fig.~\ref{dc_results}\,(a) for Reservoir 1 temperatures $T_{\rm 1} =295$ mK, at phonon bath temperature $T_{\rm 0} = 80$ mK. The close-up of the main feature is plotted in panel (b) of Fig.~\ref{dc_results}, for the Reservoir 1 temperatures $T_{\rm 1} =295, 270, 235$ and $70$ mK. 

The main features of the observed power on Reservoir 2 are as follows: (i) There is an absorption peak at half-integer flux positions (in units of $\Phi_0$) of magnitude $P\sim 10$ fW, which is more than an order of magnitude higher than in that observed earlier in a transmon system~\cite{Ronzani2019}, while varying the magnetic field (see panel (c) of Fig.~\ref{dc_results}). (ii) The central peak is accompanied by two main satellite peaks, positioned at flux values where the qubit energy is degenerate with that of the resonators (see also Fig.~\ref{working concept}). (iii) Under a more careful measurement (Fig.~\ref{dc_results}\,(d)), we observe additional weaker side peaks outside the central triplet. \textcolor{black}{(iv) When lowering the temperature of the Reservoir 1 below the bath temperature \cite{Nahum, PekolaJP}, we observe cooling of Reservoir 2, as expected (see the lowest trace in Fig.~\ref{dc_results}\,(b)). Electronic cooling is achieved by voltage biasing the NIS junction at the optimal cooling point, $\Delta/e$, allowing only high-energy (hot) electrons in the normal metal to tunnel into the superconductor, while the lower-energy electrons remain in the normal metal. As a consequence of the selective evacuation of high-energy electrons from the normal metal, electronic cooling is observed~ \cite{Francesco Giazotto, PekolaJP}.} (v) There is substantial `background' flux-independent heat transport between both reservoirs, which is most likely carried by the phonons (substrate).

\textcolor{black}{The theoretical model of the system, described in section V, can account for the observations (i)-(iv), see panels (e, f) of Fig.~\ref{dc_results}. The central peak at $\Phi/\Phi_0 =0.5$ arises from the minimum energy of the qubit allowing for thermal transitions (excitations) from the ground to the excited state at this position and at sufficiently high temperature of Reservoir 1. While moving away from this minimum energy position (see Fig.~\ref{spec_results}\,(b)), these excitations quickly diminish, and the power drops back to the base level. But when the qubit becomes degenerate with the resonators, this opens a new transport channel in the hybridized qubit-resonator system leading to the side peaks in the triplet.} \textcolor{black}{Due to the multiple modes of the resonator there are several satellite peaks around the sweet spot, as shown in panel (d) of Fig. \ref{dc_results}.}

\section{Qubit read-out}
\vspace{2mm}

\begin{figure}[ht]
\centering
\includegraphics [width=0.99\columnwidth] {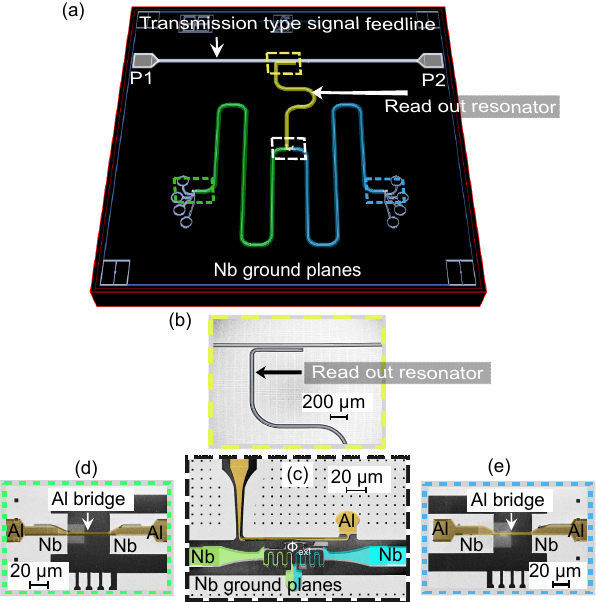}
\caption{\textcolor{black}{Qubit spectroscopy device. (a) Device 2: Design layout of the measured spectroscopy device. Colored semi-solid boxes represent different components of the device shown in panels b--e. (b) SEM image showing capacitive coupling of the read-out resonator to the transmission feed-line. (c) Electron micrograph of the flux qubit. The narrow part of the read-out resonator is made of Al. (d),(e) In Device 2, superconducting Al bridges are utilized to link the Resonator 1 and Resonator 2 to the ground plane. The SEM images have been colored to enhance visibility of the features.}}
\label{spec_device}
\end{figure}

\textcolor{black}{Here, we report on Device 2 that enables us to perform rf spectroscopy on it to extract parameters such as the resonator frequency, qubit transition frequency and their coupling strengths. 
Since the resonator frequency and coupling strength are merely geometry dependent quantities, the spectroscopy replica gives estimates for these parameters. Yet the qubit frequency is more susceptible to precise fabrication processes, and therefore variations between Device 1 and 2 may be large.} These parameters are essential for supporting the interpretation of the thermal transport measurements performed using Device~1. The Device 2 design schematic is shown in Fig. \ref{spec_device}. \textcolor{black}{It features a signal feed-line that is capacitively coupled to one end of a quarter-wavelength read-out resonator, with a designed read-out frequency $f_{read-out}=$ $7.0~$GHz, see panel (b) of Fig. \ref{spec_device}.}  The opposite end of the read-out resonator is positioned in close proximity to the flux qubit and is inductively terminated to the on-chip ground plane, as shown in Fig. \ref{spec_device} (c). As reported in Ref.~\cite{Randy}, the broadening of the resonance line-width is expected with copper terminations. Therefore, for measurement convenience, in the spectroscopy device, copper terminations are not used. Instead, a superconducting aluminum bridge is employed to connect the resonator to the niobium ground plane, see panels (d) and (e) of Fig. \ref{spec_device}. We use continuous wave transmission coefficient measurements to probe the device characteristics.

Details about the cryostat and the measurement setup are covered in Appendix B of this article and they remain similar to the scheme employed in Ref. \cite{R.Upadhyay}. To characterize the qubit properties we perform single and two-tone spectroscopy. The weakly coupled qubit-read-out resonator interaction is detected by the change in the read-out resonator frequency as a function of applied magnetic field. The qubit energy spectra as a function of applied magnetic flux is measured using two-tone spectroscopy where the second-tone is used to excite the qubit and the probe tone detects the change in qubit state. For more details on single and two-tone spectroscopy techniques see Ref. \cite{R.Upadhyay}. \textcolor{black}{In this article, the two-tone qubit spectroscopy data is presented in panels (a) and (b) of Fig. \ref{spec_results}. The splitting visible in panel (a) highlights the strong interaction $(2g)$ between the qubit frequency and the resonances observed in the device, while panel (b) reveals the qubit energy spectrum, which eventually reaches a minimum at approximately 3.9 GHz at qubits's degeneracy point.} \textcolor{black}{To estimate the qubit-resonator interaction energy and qubit's persistent current $(I_{p})$ of the device we fit the plots shown in panel (a) and (b) of Fig. \ref{spec_results}, with well-known Jaynes-Cummings model (J-C) \cite{JC} and $\hbar \, \epsilon =  2\,I_{p}\,(\,\Phi\,-\,\Phi_{0}/2\,)$ \cite{Niskanen}, where $\omega=\sqrt{\epsilon^2+\Delta^2}$. In particular, we may write $hf_q(\Phi) = hf_q^0\sqrt{1+\left(\frac{I_p}{hf_q^0}\right)^2\left(\frac{\Phi}{\Phi_0}-\frac{1}{2}\right)^2}$, 
where $f_q(\Phi)$ and $f_q^0$ represent the qubit frequencies at flux $\Phi$ and at $\Phi = 0$, respectively. For this particular sample, we obtain $I_p = 21~\mathrm{nA}$.}

In this section, we analyze the measured system theoretically to explain the origin of the observed heat transport characteristics of the device. The total Hamiltonian $\hat{\mathcal H}$ for the setup shown in Fig.~\ref{e-pg_coupling}\,(b), and derived in Appendix E, excluding the Reservoir 1 and Reservoir 2, is given by
\begin{equation}\label{Hamiltonian-1}
	\hat{\mathcal H}=\hat{\mathcal H}_{\rm q}+\sum_{i=1,\,2} \hat{\mathcal H}_{\rm i}+\sum_{i=1,\,2}\hat{\mathcal H}_{\rm qci}+\hat{\mathcal H}_{\rm 12}.
\end{equation} 
Here $\hat{\mathcal H}_{\rm q}=hf_{\rm q}(\Phi) \hat{b}^\dagger \hat{b}$ is the Hamiltonian of the qubit, $ \hat{\mathcal H}_{\rm i}=\sum_{i=1,\,2}hf_i \hat{a}_i^\dagger \hat{a}_i$ that of each resonator, and we approximate the coupling between the resonator $i$ and \textcolor{black}{ the qubit as $\hat{\mathcal H}_{\rm qci}=-\sum_{i=1,\,2}hg_i (\hat{a}_i+ \hat{a}_i^\dagger)(\hat{b}+ \hat{b}^\dagger)$, }and that between the two resonators as $ \hat{\mathcal H}_{\rm 12}=-h\gamma(\hat{a}_1+ \hat{a}_1^\dagger)(\hat{a}_2+ \hat{a}_2^\dagger)$~\cite{Upadhyay}. Here, $\hat{b}^\dagger$ and $\hat{b}$ represent the creation and annihilation operators for the qubit, and similarly, $\hat{a}^\dagger$ and $\hat{a}$ are for resonator $i$, $g_i$ denotes the coupling constant between the qubit and resonator $i$, and $\gamma$ between the two resonators. We use the Hilbert space $\big{\{}|0\rangle\equiv|g00\rangle$, $|1\rangle\equiv|e00\rangle$, $|2\rangle\equiv|g10\rangle$ and $|3\rangle\equiv|g01\rangle\big{\}}$, where the three entries in each state refer to the qubit which is either in the ground state $|g\rangle$ or excited state $|e\rangle$, Resonator 1, and Resonator 2, respectively. The non-vanishing transitions between the eigenstates, which contribute to the heat transport in our device, are spectroscopically examined and presented in panels (a) and (b) of Fig.~\ref{spec_results}. These transitions are modeled as illustrated in panel (c) of the same figure.

The transition rate induced by the reservoir $r~(r=1,\,2)$ between the normalized eigenstates $|\psi_i\rangle$ and $|\psi_j\rangle$ of Hamiltonian \eqref{Hamiltonian-1} reads
\begin{equation}\label{Gij-2}
	\Gamma_{i \rightarrow j}^{(r)}=\frac{1}{\hbar^2}|\langle \psi_i|\frac{\partial \hat{\mathcal H}}{\partial \Phi}|\psi_j\rangle|^2 S_\Phi^{(r)}(\omega_{ij}),
\end{equation}
where $|\psi_k\rangle=\sum_{i=0}^{3} a_{ki} |i\rangle$, with the coefficients $a_{ki}$ provided in Appendix 3. Here, $S_\Phi (\omega_{ij})$ represents the spectral density of flux noise, which will be discussed later in this section. Further details on $\partial \hat{\mathcal H}/{\partial \Phi}$ and the matrix elements $|\langle \psi_i|{\partial \hat{\mathcal H}}/{\partial \Phi}|\psi_j\rangle|^2$ can be found in Appendix 3 of the article. 

In order to obtain the noise spectrum $S_\Phi^{(r)}(\omega_{ij})$ in the expression of the transition rates, $S_\Phi^{(r)}(\omega_{ij})$ in Eq.~\eqref{Gij-2}, we consider the influence of each (mutually uncorrelated) reservoir separately, and in the following we therefore drop the reservoir indices in the corresponding quantities. We first note that the flux $\Phi(t)$ and current $I(t)$ through the input coil of inductance $M$ of the flux qubit are related as $\Phi(t)= M\,I(t)$ for a directly coupled qubit as ours (see Fig.~\ref*{e-pg_coupling}\,(c)). Then for the spectral density of noise we have $S_\Phi(\omega_{ij})=M^2\,S_{I}(\omega_{ij})$. Due to the coupling via a $\lambda/2$ coplanar wave resonator with resonance frequency $f_r$, we have the flux noise on the qubit at frequency $f=\omega/2\pi$ in form $S_\Phi(\omega)=M^2 |t|^2 S_{I_n}(\omega)$, where $|t|^2$ is the transmission through the resonator reading
\begin{equation}\label{transmission}
	|t|^2=\frac{1}{\cos^2(\pi\frac{f}{f_{\rm r}})+\big{(}\frac{Z_\infty}{R}\big{)}^2\sin^2(\pi\frac{f}{f_{\rm r}})}.
\end{equation}
Here $Z_\infty\approx 50\,\Omega$ is the designed impedance of the transmission line. The bare current noise spectrum of the resistor $R$ at temperature $T=1/(k_B\beta)$ is $S_{I_n}(\omega)=2\hbar\omega/\big{[}R(1-e^{-\beta\hbar\omega})\big{]}.$ The diagonal elements of the density matrix under non-driven conditions are given by the steady-state solution of
\begin{equation}\label{ss-ME}
	\sum_{j}\Gamma_{j \rightarrow i}\rho_{jj}-\sum_{j}\Gamma_{i \rightarrow j}\rho_{ii}=0.
\end{equation}
Here $\Gamma_{i\rightarrow j}=\Gamma_{i\rightarrow j}^{r=1}+\Gamma_{i\rightarrow j}^{r=2}$. The power to reservoir $r$, ($r=1,\,2$) then reads
\begin{equation}\label{P-gen}
	P_{1\rightarrow 2}=\sum_{k,l} \rho_{k}\hbar\omega_{kl}\Gamma_{k\rightarrow l}^{(r=2)}.
\end{equation}
Figures \ref{dc_results}\,(a-d) and \ref{dc_results}\,(f) show the flux dependence of the power in the experiment and that obtained by the model above, respectively. The parameters used in the calculations and plotting the theoretical curves are as follows: $R= 6\,\Omega$, $L=0.8\,$nH, $I_p=30\,$nA, the minimum energy of the qubit at half flux, $f_{\rm q}^{(0)}=2\,$GHz , and $f_{\rm r}=7\,$GHz. \textcolor{black}{Our model has some simplifying assumptions. We restrict it to a single-excitation basis and use the Fermi golden rule transition rates. These formulations are fine as long as temperatures are sufficiently low and couplings are not too strong.}

\textcolor{black}{The spectroscopically observed values for Device~2 are $f_{r} \approx 6.4~\mathrm{GHz}$, $f_{q} \approx 4.0~\mathrm{GHz}$, and $I_{p} \approx 21~\mathrm{nA}$. These parameters, although only approximate, support the theoretical modeling used to reproduce the experiments performed with Device~1; although they exhibit modest but noticeable deviations from the modeled values. These deviations may be attributed to the contributing factors outlined below. Architecturally, Device 1 adopts an N–resonator–flux qubit–resonator–N configuration, whereas Device 2 follows an S–resonator–flux qubit–resonator–S layout, substituting the normal-metal terminal with a Al strip to facilitate spectroscopy. Both devices share fabrication steps up to qubit integration, after which Device~1 undergoes additional processing to incorporate heat reservoirs and NIS junctions, which may alter the Josephson junction characteristics. Furthermore, the read-out resonator present in Device~2 (Fig.~\ref{spec_device} (c)), may influence the electromagnetic environment and contribute to frequency deviations. Finally, it is well known that precise frequency targeting in flux qubits with three Josephson junctions is experimentally challenging, particularly in research-grade fabrication environments.}


\begin{figure}[ht]
\centering
\includegraphics [width =0.78\columnwidth] {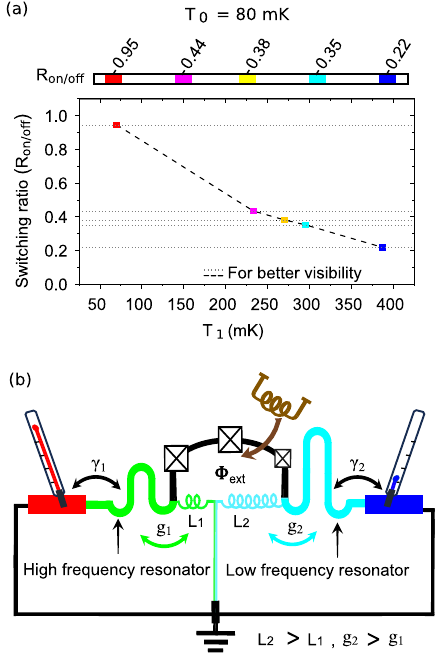}
\caption{Potential applications of the flux-qubit  device. (a) Switching ratio from Eq. (\ref{Performance}) vs Reservoir 1 temperatures at $T_{\text{0}} = 80$ mK, with semi-solid line and horizontal grid lines for visibility, demonstrates a close to $100 \%$ switching ratio. \textcolor{black}{The color bar corresponds to the values of $R_{on/off}$ at different temperatures of Reservoir 1 ($T_{1}$). (b) Towards quantum heat diode and Otto refrigerator: Asymmetry introduced by tailoring resonator frequencies and qubit coupling.}}
\label{dc_valve_perform}
\end{figure}

\vspace{5mm}
\section {Conclusion and Outlook}
\label{Outlook}
\vspace{2mm}

A flux qubit presented here offers several advantages for quantum thermodynamics experiments and devices as compared to the common transmon qubit. \textcolor{black}{The main merits are the anharmonicity which remains robust against strong coupling to the resonators and reservoirs and the galvanic coupling of the qubit rendering efficient transport.} \textcolor{black}{Notably, the inductive coupling design in the reported device architecture facilitates access to the ultra-strong coupling regime by adjusting the length-dependent inductance of the coupling element ~\cite{R.Upadhyay}. This leads to the ability to transfer large heat currents via the qubit (proportional to $g^2$ for coupling strength $g$ based on Fermi golden rule arguments) without significant compromises; an order of magnitude enhancement was achieved here as compared to earlier transmon experiments \cite{Ronzani2019}. Secondly, weak non-reciprocity in heat transport has been demonstrated using a transmon qubit earlier, but this is limited by the nearly equidistant energy levels, i.e., transmon is almost linear.} A flux qubit would thus present a two-level qubit (maximal non-linearity), with strong directionality of heat transport, exhibiting an efficient thermal diode \cite{Upadhyay}. Additionally, due to the sharply peaked heat transport in the femtowatt range around $\Phi=\Phi_0/2$, as demonstrated by our experiment, the system operates as a heat switch controlled by magnetic flux. Measured at the temperature of 80 mK, the on/off switching ratio ${R}_{\rm on/off}$ of the heat current, approaches $100\%$, as estimated using the ratio of the corresponding powers as
\begin{equation}	\label{Performance}
{R}_{\rm on/off}= \frac{P_{\rm on} -P_{\rm off}}{P_{\rm on}},
\end{equation} 
and shown in panel Fig.~\ref{dc_valve_perform}\,(a). Here the on and off conditions are defined as those at $\Phi=\Phi_0/2$ and $\Phi=0$, respectively. \textcolor{black}{When $ T_1 $ and $ T_2 $ are low, with $ T_1 $ even less than $ T_2 = 80 \, \text{mK} $, which corresponds to the maximum cooling point of the NIS control junction, the magnitude of phonon heat transport is weakest, and photon transport dominates in system. As a result, we approach a switching ratio close to 100\%. However, when $ T_1 $ is increased to 230\,\text{mK} and above, the phonon contribution becomes dominant (since phonon heat conductance generally obeys a $ T^3 $ dependence \cite{Pobell}), and the switching ratio, determined by the overall power including both photons and phonons, diminishes.}

Based on all the merits listed above, a flux qubit enables the development of efficient quantum thermal machines. Introducing asymmetry to the system in terms of the frequencies of the two superconducting resonators and tailoring qubit coupling inductances, one finds opportunities for developing not only quantum heat diodes \cite{Jorden, Upadhyay} but also quantum refrigerators \cite{B.Karimi}. The required  asymmetries are achieved by choosing the length of the corresponding element, resonator or coupling inductance, as shown in Fig. \ref{dc_valve_perform} (b). This refrigerator can pump heat under the application of alternating magnetic flux to the qubit, from a cold reservoir via the lower frequency resonator to the hot reservoir via the higher frequency one.
 
\vspace{5mm}

\textbf{Acknowledgment} 
We thank Luca Magazzu, Milena Grifoni, and Elisabetta Paladino for useful discussions. We acknowledge the financial support from the Research Council of Finland grants (grant number 297240, 312057, 303677, and 349601 (THEPOW)), and from the European Union’s Horizon 2020 research and innovation programme under the European Research Council (ERC) programme (grant number 742559) and Marie Sklodowska-Curie actions (grant agreements 766025).We sincerely recognize the provision of facilities by Micronova Nanofabrication Centre, and OtaNano - Low Temperature Laboratory of Aalto University which is a part of European Microkelvin Platform EMP (grant number. 824109), to perform this research. We thank and acknowledge VTT Technical Research Center for provision of high quality sputtered Nb films.

\vspace{5mm}

\textbf{Author contributions}
\newline
The device design and fabrication was done by R.U. The measurements were carried by R.U, Y-C.C and D.S. The data analysis is done by R.U, B.K, and C.D.S. The theoretical model was conceived by J.P.P and B.K. The technical support was provided by D.S, Y-C.C. and J.T.P. The manuscript was written by R.U., J.P.P and B.K., with important contributions from all other authors. The work was supported and supervised by J.P.P. 

\vspace{5mm}

\textbf{Data availability}
\newline
The data that support the plots within this article are available from the corresponding author upon reasonable request.
\newline

\vspace{5mm}

\textbf{Competing interests}
\newline
The authors declare no competing interests.
\newline

\vspace{5mm}

\section*{Appendix A: Device fabrication}
\vspace{5mm}

\begin{figure*}[ht]
\centering
\includegraphics [width = 0.9\textwidth] {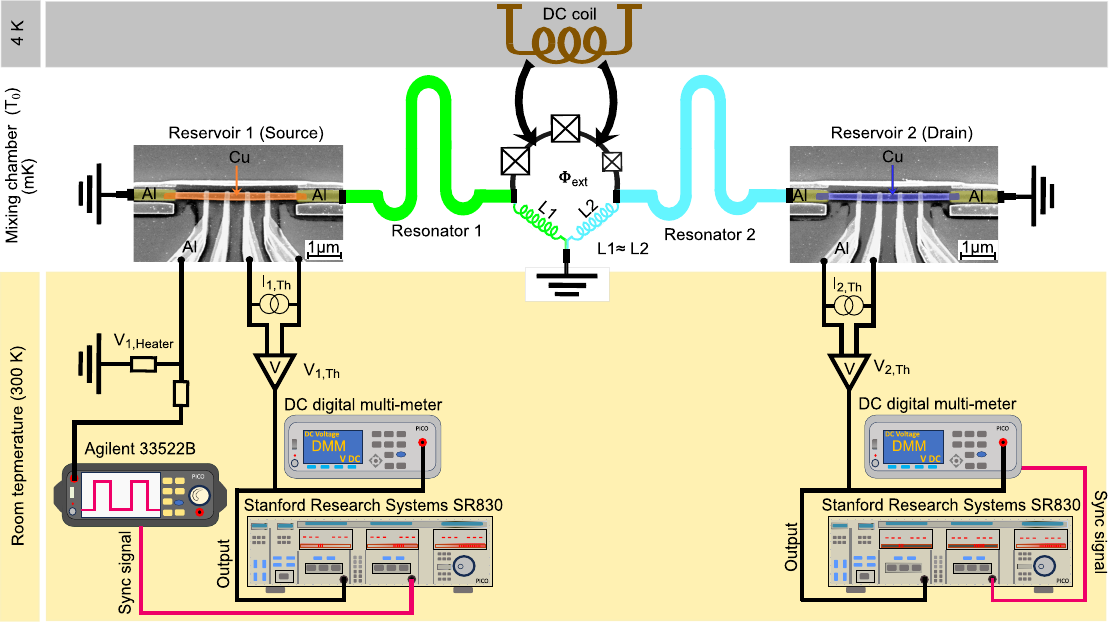}
\caption{ \textcolor{black}{Thermal conductance measurement setup of the reported device at various temperature stages. The top row, shown in gray, represents the 4K stage where the global magnetic coil is placed. Below it, the middle row corresponds to the mixing chamber of the cryostat, at temperature $T_0$, where the device is located. The third and bottom-most row, highlighted in yellow, shows the configuration of the electronic equipments used in the experiments, which are situated at room temperature.}}
\label{T1_T2_DC}
\end{figure*}
\vspace{0.2cm}

A 675 $\mu$m thick highly resistive Si wafer is used as a substrate. On top of it, a 30~nm thick dielectric film of Al$_2$O$_3$ is deposited using atomic layer deposition (ALD), followed by a 200~nm thick deposition of niobium metal using DC magnetron sputtering. The resonator mask is patterned over a uniform positive e-beam resist using e-beam lithography. After development of patterns, the wafer is baked for a few minutes to strengthen the remaining resist and secure the main structures, avoiding their over-etching. Reactive-ion-etching (RIE) is performed using $SF_{6}$ chemistry to etch the exposed Nb metal. In the last step, the remaining resist is removed using $O_{2}$ descum dry resist stripping.

\begin{figure*}[ht]
\centering
\includegraphics [width = 0.7\textwidth] {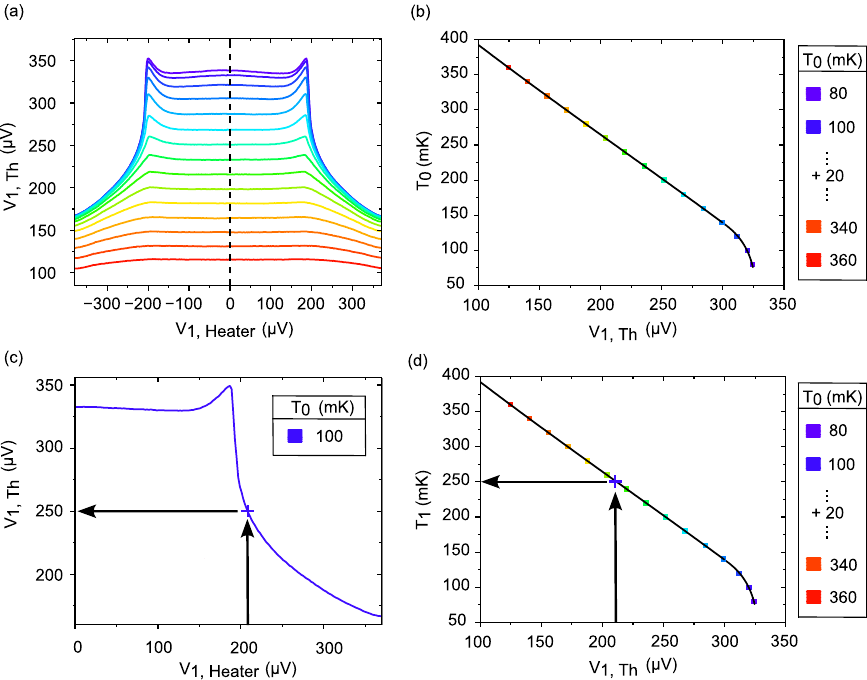}
\caption{Thermal voltage to temperature conversion. (a). Typical thermal voltage curve of the SINIS thermometers when biased with varying heating voltages at different cryostat temperatures. (b). Reservoir 1 thermal voltage ($V_{1,\text{Heater}}$), plotted against various cryostat temperatures, at $V_{1,\text{Heater}}=0$, indicated by the dashed line in panel (a). (c-d). The SINIS thermal voltage curve at a cryostat temperature of 100 mK indicates that a voltage of $\approx$  210 $\mu$V corresponds to a temperature of 250 mK. Black solid arrows are used for better visibility.}
\label{Conversion to Temp}
\end{figure*}

For patterning the qubit, a bi-layer resist stack of P(MMA-MMA) copolymer and PMMA is spin-coated, followed by an e-beam exposure. The bi-layer stack is used to create an undercut useful for multi-angle evaporation. We use a solution of Methyl-Isobutyl-Ketone:Isopropanol alcohol (MIBK:IPA volume ratio, 1:3) to develop the PMMA top layer. The undercut profile is provided by a second development step using a Methyl-glycol-Methanol solution with a volume ratio of 1:2. A high-vacuum (HV) e-beam evaporator with an in-situ plasma etching module is used for evaporation. Before evaporation, the sample is exposed to an argon beam for about 40 seconds to clean possible deposited native oxide, enabling a clean contact at Nb-Al interfaces. Using the Dolan bridge shadow evaporation technique~\cite{Dolan}, a 30~nm thick Al is deposited at $+18^\circ$. After the first layer deposition, the devices are oxidized using an in-situ oxidation module (4.3~mbar, for 4 minutes), followed by another Al layer deposition at a second angle ($-18^\circ$). Finally, the devices are submerged into a hot $+52^\circ$C acetone bath for about an hour to perform the metal lift-off.

For fabricating heat reservoirs and thermometers, which is the next step, we use the three-angle evaporation method. After spin-coating another bi-layer resist stack of P(MMA-MMA) copolymer and PMMA, e-beam exposure, development, and native oxide milling, a 20~nm thick Al layer is deposited at $+39^\circ$, forming the superconducting (S) part of the NIS structures. The deposited layer is oxidized under 2.5~mbar chamber pressure for about 5 minutes, forming an insulating tunnel oxide barrier (I). Then a 2~nm thick Al layer is deposited at $-6.5^\circ$, followed by the deposition of a 50~nm thick normal metal Cu (N) at the same tilt angle. The 2~nm Al layer makes the adhesion of the Cu layer stronger to the substrate. In the last evaporation step, a 90~nm thick layer of Al is deposited at $+20^\circ$. This layer enables a clean contact between the NIS structures and the superconducting resonators. 

The device is analyzed under SEM, followed by dicing and bonding for cryogenic measurements. Integrated within the same device, we have a replica of our flux qubit connected to two pads, allowing Al bonding wires to measure the normal state resistance of the SQUID loop. The measured resistance of the replica is $\approx 3.4~{\rm k}\Omega$ at liquid He temperatures.

\vspace{2mm}
\section*{Appendix B: Measurements setup}
\setcounter{subsection}{0}
\renewcommand{\thesubsection}{\arabic{subsection}}

\subsection{Thermal Measurement setup}
\vspace{1mm}

The sample is glued and wire-bonded ultrasonically to a brass sample stage. Encapsulated within two brass caps acting as radiation shields, the sample holder is then plugged to the mixing chamber (MXC) of a custom-made plastic wet dilution refrigerator with base temperature $\approx 70~$mK. For read-out purpose, we use conventional thermocoax-filtered cryogenic lines that connect the sample stage from MXC to the room-temperature break-out box. For low-impedance loads, the effective signal bandwidth for the cryogenic filtered lines are between 0-10~kHz. For measurements, programmable voltage sources and Keysight digital multi-meters were employed. For amplification of output voltages we use a room temperature Femto DLPA-100-F-D voltage amplifier. Agilent 33522B waveform generator is used as the bais voltage source. For reading the heat current modulation as a function of applied magnetic field we use SRS SR830 DSP lock-in amplifier which is synchronized to the square-wave modulation of the voltage bias from the Agilent 33522B waveform generator. We perform measurements at several MXC temperatures. The MXC temperatures are controlled by applying voltage to a resistor which is connected to the MXC. For accurate analysis, the calibration of the integrated thermometers (realized as SINIS configuration) were performed at various cryostat temperatures, ranging from $80~$mK up to $360~$mK. This is done by measuring the voltage drop across the SINIS configuration (set at a current bias of $I_{th} = 50~$pA), for each cryostat temperature, with no heater bias voltage applied ($V_{1, Heater} = 0$V). \textcolor{black}{After setting the MXC to the next higher desired temperature, we wait for over 20 minutes before performing any measurements. This ensures that the device is effectively thermalized to the MXC, allowing us to assume that the phonon temperature is in equilibrium with the MXC temperature. The MXC temperature is monitored using a ruthenium oxide thermometer, calibrated against a Coulomb blockade thermometer. For measurements involving the modulation of the voltage drop across the reservoir thermometers (heat current) as a function of the applied magnetic field, the procedure is as follows: Heat-current measurements as a function of magnetic flux are initiated by stabilizing at the initial flux point for 5 seconds. After this, at each subsequent flux point the system is allowed to stabilize for 2.1 seconds before data acquisition. As previously described, this process is carried out at various values of $T_0$, with a waiting time of over 20 minutes after the cryostat temperature is increased to ensure proper thermalization.}

\renewcommand{\thesubsection}{\arabic{subsection}}
\subsection{Spectroscopy measurement setup}

The spectroscopy device, which includes a signal feed line, read-out resonator, and a qubit galvanically coupled to both resonators (as shown in Fig. \ref{spec_device}), is mounted on a gold-plated copper sample stage and placed into a dilution refrigerator with a base temperature of $\approx15~$mK. Similar to reported measurement scheme in Ref. \cite{R.Upadhyay}, the sample holder is equipped with an external superconducting coil for the application of a dc-magnetic field. Suppression of black-body radiation is achieved by employing a series of impedance-matched cryogenic attenuators placed at different temperature stages inside the cryostat. Once attenuated, the input signal passes through the device and is directed through two isolators situated in the cryostat mixing chamber. \textcolor{black}{Subsequently, it undergoes amplification by a $42~$dB low-noise high-electron-mobility transistor (HEMT)} amplifier at the 4-K stage, followed by further amplification of $52~$dB using two additional amplifiers positioned at room temperature. For single-tone spectroscopy, we utilize a commercially available vector network analyzer (VNA). For two-tone spectroscopy, we use a commercial microwave signal generator. The frequency range of interest, 4-8 GHz, is supported by the setup. The measurement setup and procedures for single-tone and two-tone spectroscopy are similar to those outlined in Ref. \cite{R.Upadhyay}.

\section*{Appendix C: Thermal transport experiments}

\setcounter{subsection}{0} 
\renewcommand{\thesubsection}{\arabic{subsection}}
\subsection{Thermal Measurements}

A custom made, plastic wet dilution refrigerator is used for thermal conductance measurements. These measurements are conducted over a mixing chamber (MXC) temperature range of $80$ to $360~$mK, which is controlled by applying voltage to a resistor located at MXC of the cryostat. The device achieves effective thermalization with the MXC, allowing us to assume that the phonon temperature, $T_{\rm 0}$, is in equilibrium with that of the MXC. This temperature is accurately measured using a ruthenium oxide thermometer, which has been calibrated with a Coulomb blockade thermometer for precision. The detailed measurement setup configuration is provided in Appendix B. Reservoir 1 is locally heated using an NIS junction which is biased by an external DC voltage source $V_{\rm{ 1,\,Heater}}$. The Reservoir 1 and Reservoir 2 thermometers are biased with a small current of $I_{th} = 50~$pA, and the voltage drop across the thermometers ($V_{\rm{ 1,\,Th}}$ and $V_{\rm{ 2,\,Th}}$, respectively), is recorded at several cryostat temperatures $T_0$. To determine the temperatures of the Reservoir 1 ($T_{\text{1}}$) and Reservoir 2 ($T_{\text{2}}$), we measure the thermal voltage drop across both thermometers (set at a current bias of $I_{th} = 50~$pA), for each cryostat temperature ($T_{\text{0}}$), with no heating bias voltage applied ($V_{\rm{ 1,\,Heater}}=0$). The recorded thermal voltage is converted to its respective temperature using the calibration method described in Appendix B. Using the calibrated data, the absorbed power by Reservoir 2, $P_{1 \to 2}$ ,is then calculated using expression, $P_{\rm ep,2}= \Sigma \mathcal{V}\,(T_{\rm 2}^{5} - T_{\rm 0}^{5})$. The magnetic flux controlling the energy of the qubit is provided by a external superconducting coil located at 4~K temperature. To observe the modulation in the voltage drop of thermometers, $V_{\rm{ 2,\,Th}}$, at Reservoir 2, we heat the Reservoir 1 ($V_{\rm{ 1,\,Heater}}\neq0$) simultaneously sweeping the magnetic field, and record the change in $V_{\rm{ 2,\,Th}}$. After calibrating the data, we arrive at the figure presented in Fig.~\ref{dc_results}.

\begin{figure*}[ht]
\centering
\includegraphics [width = 0.78\textwidth] {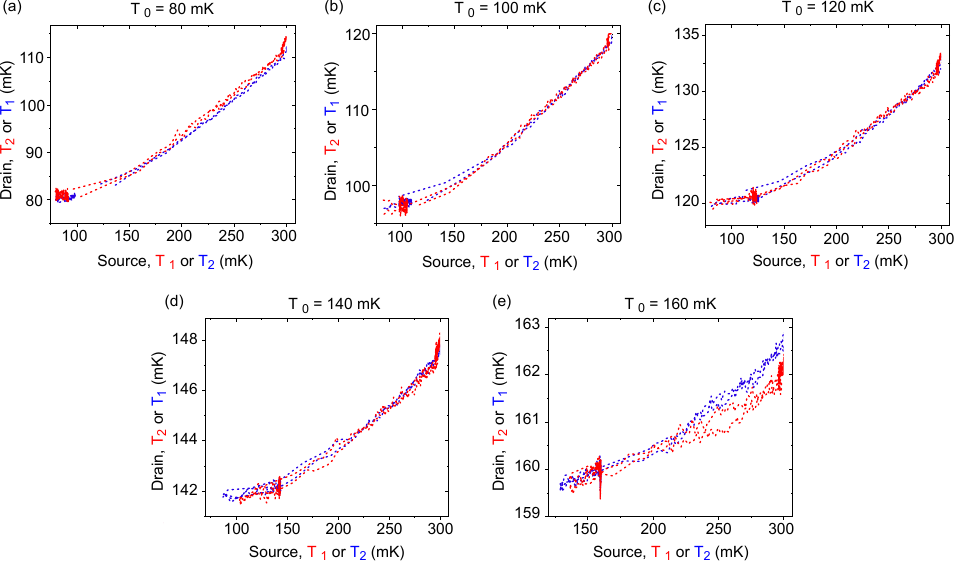}
\caption{\textcolor{black}{The ideal symmetric device. (a)  Correlation between the temperatures of both reservoirs at different phonon temperatures. The x-axis represents the temperature of the heated reservoir (source), which is controlled using an NIS junction, while the y-axis shows the temperature of the drain reservoir. Each panel corresponds to a different phonon temperature $T_{0}$: (a) 80~mK, (b) 100~mK, (c) 120~mK, (d) 140~mK, and (e) 160~mK.}}
\label{T1_T2_DC}
\end{figure*}

\renewcommand{\thesubsection}{\arabic{subsection}}
\subsection{Thermal voltage to Power conversion: Mechanism} 

Figure~\ref{Conversion to Temp}\,(a) presents the measurement of the thermometer voltage, $V_{\rm{ 1,\,Th}}$, at a fixed applied current of $ I_{\rm{1, Th}}=50$~pA for various bath temperatures, $T_0=80 - 360$~mK, with a temperature interval of 20 mK between each measurement, from top to bottom, against the voltage, $V_{\rm{ 1,\,Heater}}$, across the NIS heater. The equilibrium values of $V_{\rm{ 1,\,Th}}$ at $V_{\rm{ 1,\,Heater}}=0$, represented by symbols in Fig.~\ref{Conversion to Temp}\,(b) (with the position indicated by the black dashed line in panel (a)), are plotted as a function of bath temperature, $T_0$. These data points $V_{\rm{ 1,\,Th}}( V_{\rm{ 1,\,Heater}}=0)$ serve as the calibration for the corresponding reservoir, here the Reservoir 1 temperature. The solid black line represents a fit to the measured data. At $80$~mK $\leq T_1 < 135$~mK, the fitting function is $T_1=a+b*\lg (c-V_{\rm{ 1,\,Th}})$ with $a$, $b$ and $c$ as constant \cite{Karimi B.}, and at 135 mK $\leq T_1 < 370$ mK we used standard third-order polynomial (nearly linear) function. Panels \ref{Conversion to Temp}\,(c) and \ref{Conversion to Temp}\,(d) display how the Reservoir 1 temperature $T_1$ is extracted. With the same procedure we can obtain the Reservoir 2 temperature, $T_2$. The measured temperatures are then converted to power using Eq.~(1) described in the main text to obtain the power transmitted to the Reservoir 2.

\renewcommand{\thesubsection}{\arabic{subsection}}
\subsection{The ideal symmetric device}

The architecture of the reported device is symmetric which is also reflected by thermal properties of the device. After calibrating the thermal voltage curves of both reservoirs islands and converting them to their respective temperatures, the data for both reservoirs, expressed in temperature, are plotted against each other. The plot is reported in Fig. \ref{T1_T2_DC}. Panels (a), (b), (c), (d) and (e) in Fig. \ref{T1_T2_DC} corresponds to the phonon temperatures of $80$, $100$, $120$, $140$, and $160$~mK. In Fig. \ref{T1_T2_DC}, we can observe that the temperature curves for both Reservoir 1 and Reservoir 2 are practically identical as expected from the symmetry of the device. In panel (e) of Fig. \ref{T1_T2_DC}, a slight difference becomes noticeable when the Source Reservoir temperature exceeds 200 mK. However, the difference between the two curves remains below 1 mK, which is remarkable.

\clearpage
\onecolumngrid

\section*{Appendix E: Theoretical model}

\setcounter{subsection}{0} 
\renewcommand{\thesubsection}{\arabic{subsection}}

\subsection{Flux qubit hybridized with two resonators coupled to two reservoirs}

\vspace{0.2cm}
\begin{figure}[h!]
	\centering
	\includegraphics [width=0.6\textwidth] {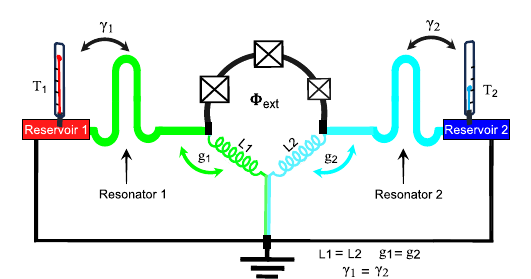}
	\caption{Device circuit schematic. A system of three junction flux qubit coupled with two resonators, which are connected to two separate reservoirs.
	\label{Supp_design}}
\end{figure}

The total Hamiltonian $\hat H$ for the setup shown in Fig.~\ref{Supp_design}, excluding the resistors to the left and right, is given by
\begin{equation}\label{Hamiltonian-1}
	\hat{\mathcal H}=\hat{\mathcal H}_{\rm q}+\sum_{i=1,\,2} \hat{\mathcal H}_{ i}+\sum_{i=1,\,2}\hat{\mathcal H}_{{\rm c}i}+\hat{\mathcal H}_{\rm 12},
\end{equation} 
where $\hat{\mathcal H}_{\rm q}$ is the Hamiltonian of the qubit, $ \hat{\mathcal H}_{i}$ that of each resonator, $ \hat{\mathcal H}_{{\rm c}i}$ the coupling between the resonator $i$ and the qubit, and $ \hat{\mathcal H}_{\rm 12}$ the coupling between the two resonators. We write Eq.~\eqref{Hamiltonian-1} in the form~\cite{Upadhyay}
\begin{equation}\label{Hamiltonian-2}
	\hat{\mathcal H}=hf_{\rm q}(\Phi) \hat{b}^\dagger \hat{b}+\sum_{i=1,\,2}hf_i \hat{a}_i^\dagger \hat{a}_i-\sum_{i=1,\,2}hg_j (\hat{a}_j+ \hat{a}_j^\dagger)(\hat{b}+ \hat{b}^\dagger)-h\gamma(\hat{a}_1+ \hat{a}_1^\dagger)(\hat{a}_2+ \hat{a}_2^\dagger).
\end{equation}
Here, $\hat{b}^\dagger$ and $\hat{b}$ represent the creation and annihilation operators for the qubit, and similarly, $\hat{a}^\dagger_i$ and $\hat{a}_i$ are for resonator $i$, $g_i$ denotes the coupling constant between the qubit and resonator $i$, and $\gamma$ between the two resonators. $\hat{\mathcal H}_{\rm q}$ can be written as
\begin{equation}\label{Hq}
	\hat{\mathcal H}_{\rm q}=hf_{\rm q}(\Phi) |e\rangle \langle e|.
\end{equation}
In the four-level basis of $\big{\{}|0\rangle\equiv|g00\rangle$, $|1\rangle\equiv|e00\rangle$, $|2\rangle\equiv|g10\rangle$ and $|3\rangle\equiv|g01\rangle\big{\}}$, where the entries in each state refer to the qubit which is either in the ground state $|g\rangle$ or excited state $|e\rangle$, left resonator, and the right resonator, respectively, the matrix form of the Hamiltonian is given by
\begin{eqnarray} \label{hamilto-matrix}
\mathcal H= h\left(
	\begin{array}{cccccc}
		0& 0 &0& 0\\ 0 & f_{\rm q} &-g_1 & -g_2\\ 0 & -g_1^*&f_1&\gamma\\ 0 &-g_2^*& \gamma &f_2 
	\end{array}
	\right).
\end{eqnarray}
We assume that the frequencies of the two resonators and their couplings to the qubit are equal, i.e., $f_1=f_2\equiv f_{\rm r}$, and $g_1=g_2\equiv g$. The eigenenergies $\lambda_k$ of this Hamiltonian are 
\begin{eqnarray}\label{eigenE}
	&&E_1=0\nonumber\\&&
E_2=hf_{\rm r}\nonumber\\&&
E_3=\frac{h}{2}[f_{\rm q}+ f_{\rm r}-\sqrt{(f_{\rm q}-f_{\rm r})^2 + 8 g^2}]\nonumber\\&&
E_4=\frac{h}{2}[f_{\rm q}+ f_{\rm r}+\sqrt{(f_{\rm q}-f_{\rm r})^2 + 8 g^2}].
\end{eqnarray}
The normalized eigenstates corresponding to these eigenenergies are:
\begin{eqnarray} \label{es01}
	|\psi_0\rangle = \left(
	\begin{array}{cccc}
		1 \\ 0 \\ 0\\ 0
	\end{array}
	\right),
	|\psi_1\rangle = \frac{1}{\sqrt{2}}\left(
	\begin{array}{cccc}
		0 \\ 0 \\-1\\ 1
	\end{array}
	\right),
\end{eqnarray}

\begin{eqnarray} \label{es01-2}
	|\psi_2\rangle = \frac{1}{\sqrt{8g^2+[-f_{\rm q}+f_{\rm r}+\sqrt{(f_{\rm q}-f_{\rm r})^2+8g^2}]^2}}\left(
	\begin{array}{cccc}
		0 \\ -f_{\rm q}+f_{\rm r}+\sqrt{(f_{\rm q}-f_{\rm r})^2+8g^2}\\2g\\ 2g
	\end{array}
	\right),
\end{eqnarray}

\begin{eqnarray}\label{es01-3}
	|\psi_3\rangle = \frac{1}{\sqrt{8g^2+[f_{\rm q}-f_{\rm r}+\sqrt{(f_{\rm q}-f_{\rm r})^2+8g^2}]^2}}\left(
	\begin{array}{cccc}
		0 \\ -f_{\rm q}+f_{\rm r}-\sqrt{(f_{\rm q}-f_{\rm r})^2+8g^2} \\ 2g\\ 2g
	\end{array}
	\right).
\end{eqnarray}

The transition rate between the states $i$ and $j$ reads
\begin{equation}\label{Gij-2}
	\Gamma_{ij}=\frac{1}{\hbar^2}|\langle \psi_i|\frac{\partial \hat{\mathcal H}}{\partial \Phi}|\psi_j\rangle|^2 S_\Phi(\omega_{ij}),
\end{equation}
where $\hbar\omega_{ij}$ is the energy separation between states $i$ and $j$.
In the Hamiltonian of Eq.~\eqref{Hamiltonian-1}, \textcolor{black}{there are two contributions, $\hat{\mathcal H}_{\rm q}$ and $\sum_{i=1,\,2}\hat{\mathcal H}_{\rm qci}$, that depend on flux $\Phi$.} The derivative of each of them with respect to flux is presented below. For $\hat{\mathcal H_{\rm q}}$ we have
\begin{equation}\label{dHqdPhi}
 \frac{\partial \hat{\mathcal H_{\rm q}}}{\partial \Phi}=h\frac{\partial f_{\rm q}(\Phi)}{\partial \Phi}\,|e\rangle \langle e|+hf_{\rm q} (\Phi)\bigg{(}\frac{\partial |e\rangle}{\partial \Phi} \langle e|+|e\rangle \frac{\partial \langle e|}{\partial \Phi}\bigg{)},
\end{equation}
This equation can be written as
\begin{equation}\label{dHqdPhi-2}
	\frac{\partial \hat{\mathcal H_{\rm q}}}{\partial \Phi}=2\,I_p\sqrt{1-\,\bigg{(}\frac{f_{\rm q}^{(0)}}{f_{\rm q}}\bigg{)}^2}\,\hat{b}^\dagger\hat{b}+\frac{f_{\rm q}^{(0)}}{f_{\rm q}}I_p\,(\hat{b}+\hat{b}^\dagger).
\end{equation}
For the coupling Hamiltonians we have
\begin{equation}\label{dHcdPhi}
	\frac{\partial \hat{\mathcal{H}}_{\rm qci}}{\partial \Phi} = -h\,g_i(\hat{a}_i+ \hat{a}_i^\dagger)\bigg{(}\frac{\partial |e\rangle}{\partial \Phi} \langle g|+|e\rangle \frac{\partial \langle g|}{\partial \Phi}+\frac{\partial |g\rangle}{\partial \Phi} \langle e|+|g\rangle \frac{\partial \langle e|}{\partial \Phi}\bigg{)},
\end{equation}
\begin{equation}\label{dHcdPhi_2}
	\frac{\partial \hat{\mathcal{H}}_{\rm qci}}{\partial \Phi}=\frac{g_i}{\Phi_0}\frac{I_p}{f_{\rm q}^{(0)}}\,\bigg{(}\frac{f_{\rm q}^{(0)}}{f_{\rm q}}\bigg{)}^2(\hat{a}_i+ \hat{a}_i^\dagger)( 2\hat{b}^\dagger\hat{b}-1),
\end{equation}
Rewriting the states $|\psi_i\rangle$ in Eqs.~\eqref{es01}, \eqref{es01-2}, and \eqref{es01-3} as
\begin{equation}
	|\psi_k\rangle=\sum_{i=0}^{3} a_{ki} |i\rangle,
\end{equation}
the non-vanishing matrix elements in the transition rates given in Eq.~\eqref{Gij-2} are as listed below
\begin{equation}\label{M10}
	M_{10}\equiv \langle \psi_1|\frac{\partial \hat{\mathcal H}_{\rm qc1}}{\partial \Phi}|\psi_0\rangle+\langle \psi_1|\frac{\partial \hat{\mathcal H}_{\rm qc2}}{\partial \Phi}|\psi_0\rangle=-a_{12} 	\frac{hg_1}{\Phi_0}\frac{I_p}{e\,f_{\rm q}^{(0)}}\bigg{(}\frac{f_{\rm q}^{(0)}}{f_{\rm q}}\bigg{)}^2-a_{13} 	\frac{hg_2}{\Phi_0}\frac{I_p}{e\,f_{\rm q}^{(0)}}\bigg{(}\frac{f_{\rm q}^{(0)}}{f_{\rm q}}\bigg{)}^2=M_{01}^*
\end{equation}
\begin{eqnarray}\label{M02}
M_{02}\equiv &&\langle \psi_0|\frac{\partial \hat{\mathcal H}_{\rm q}}{\partial \Phi}|\psi_2\rangle+\langle \psi_0|\frac{\partial \hat{\mathcal H}_{\rm qc1}}{\partial \Phi}|\psi_2\rangle+\langle \psi_0|\frac{\partial \hat{\mathcal H}_{\rm qc2}}{\partial \Phi}|\psi_2\rangle\nonumber\\&&= a_{21}\frac{f_{\rm q}^{(0)}}{f_{\rm q}}I_p-a_{22}\frac{hg_1}{\Phi_0}\frac{I_p}{e\,f_{\rm q}^{(0)}}\bigg{(}\frac{f_{\rm q}^{(0)}}{f_{\rm q}}\bigg{)}^2-a_{23} 	\frac{hg_2}{\Phi_0}\frac{I_p}{e\,f_{\rm q}^{(0)}}\bigg{(}\frac{f_{\rm q}^{(0)}}{f_{\rm q}}\bigg{)}^2=M_{20}^*
\end{eqnarray}
\begin{eqnarray}\label{M03}
	M_{03}\equiv &&\langle \psi_0|\frac{\partial \hat{\mathcal H}_{\rm q}}{\partial \Phi}|\psi_3\rangle+\langle \psi_0|\frac{\partial \hat{\mathcal H}_{\rm qc1}}{\partial \Phi}|\psi_3\rangle+\langle \psi_0|\frac{\partial \hat{\mathcal H}_{\rm qc2}}{\partial \Phi}|\psi_3\rangle\nonumber\\&&= a_{31}\frac{f_{\rm q}^{(0)}}{f_{\rm q}}I_p-a_{32}\frac{hg_1}{\Phi_0}\frac{I_p}{e\,f_{\rm q}^{(0)}}\bigg{(}\frac{f_{\rm q}^{(0)}}{f_{\rm q}}\bigg{)}^2-a_{33} 	\frac{hg_2}{\Phi_0}\frac{I_p}{e\,f_{\rm q}^{(0)}}\bigg{(}\frac{f_{\rm q}^{(0)}}{f_{\rm q}}\bigg{)}^2=M_{30}^*
\end{eqnarray}

\begin{equation}\label{M23}
	M_{23}\equiv \langle \psi_2|\frac{\partial \hat{\mathcal H}_{\rm q}}{\partial \Phi}|\psi_3\rangle= a_{21}\,a_{31}\,2§\,I_p\sqrt{1-\,\bigg{(}\frac{f_{\rm q}^{(0)}}{f_{\rm q}}\bigg{)}^2}=M_{32}^*
\end{equation}

Since the couplings between each resonator and the qubit are equal, $g_1=g_2\equiv g$, the matrix elements $M_{10}=M_{01}=0$. Then, $|\psi_1\rangle$ remains as a dark state with no transitions into or out from it. Panel (a) of Fig. 4 of the main text provides an overview of the non-vanishing transitions between eigenstates that contribute to heat transport in this setup.

\vspace{1cm}

\renewcommand{\thesubsection}{\arabic{subsection}}
\subsection{Coupling of the noise to the flux qubit}

\vspace{0.2cm}

In order to obtain the noise spectrum $S_\Phi(\omega_{ij})$ in the expression of the transition rates, $S_\Phi(\omega_{ij})$ in Eq.~\eqref{Gij-2}, we first note that the flux $\Phi(t)$ and current $I_L(t)$ through the input coil of mutual inductance $M$ of the flux qubit are related as

 \begin{figure}[h!]
	\centering
	\includegraphics  [width=0.7\textwidth] {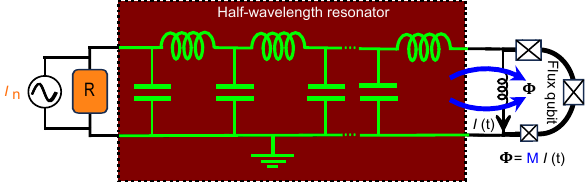}
	\caption{Circuit representation of one side of the reported device in the main text.
		\label{LC circuit}}
\end{figure}
\begin{equation}
	\Phi(t)= M\,I(t)
\end{equation}
for a directly coupled qubit as ours.
Due to the coupling via a $\lambda/2$ coplanar wave resonator with resonance frequency $f_r$, we have the flux noise on the qubit at frequency $f=\omega/2\pi$ in form 
\begin{equation}
	S_\Phi(\omega)=M^2 |t|^2 S_I(\omega),
\end{equation}
where, $S_I(\omega)$ is the bare current noise produced by the resistor, and for this distributed $LC$ resonator we have
\begin{equation}\label{transmission}
	|t|^2=\frac{1}{\cos^2(\pi\frac{f}{f_{\rm r}})+\big{(}\frac{Z_\infty}{R}\big{)}^2\sin^2(\pi\frac{f}{f_{\rm r}})},
\end{equation}
where $Z_\infty\approx 50\,\Omega$ is the impedance of the transmission line. The bare current noise spectrum of the resistor $R$ at inverse temperature $\beta$ is
\begin{equation}\label{barenoise}
	S_I(\omega)=\frac{2}{R} \frac{\hbar\omega}{1-e^{-\beta\hbar\omega}}.
\end{equation}
Based on the allowed transitions presented in Fig.~4\,(c) in the main text, the diagonal elements of the density matrix under non-driven conditions are given by
\begin{eqnarray}\label{DM}
	&&\dot{\rho}_{0}=\Gamma_{2\rightarrow 0}\rho_{2}+\Gamma_{3\rightarrow 0}\rho_{3}-(\Gamma_{0\rightarrow 2}+\Gamma_{0\rightarrow 3})\rho_{0}\nonumber\\
	&&\dot{\rho}_{1}=0\nonumber\\
	&&\dot{\rho}_{2}=\Gamma_{0\rightarrow 2}\rho_{0}+\Gamma_{3\rightarrow 2}\rho_{3}-(\Gamma_{2\rightarrow 0}+\Gamma_{2\rightarrow 3})\rho_{2}\nonumber\\
	&&\dot{\rho}_{3}=\Gamma_{0\rightarrow 3}\rho_{0}+\Gamma_{2\rightarrow 3}\rho_{2}-(\Gamma_{3\rightarrow 0}+\Gamma_{3\rightarrow 2})\rho_{3},
\end{eqnarray}
where $\Gamma_{i\rightarrow j}=\Gamma_{i\rightarrow j}^{r=1}+\Gamma_{i\rightarrow j}^{r=2}$, where the superscript $r$ refers to the both Reservoir 1 and Reservoir 2. For steady state we put $\dot{\rho}_{i}=0$, $(i=0 ... 3)$, and $\sum_{i=0}^{3} \rho_i=1$, which yields the populations of the levels ${\rho}_{i}$ as

\begin{figure}[h!]
	\centering
	\includegraphics [width=0.9\textwidth] {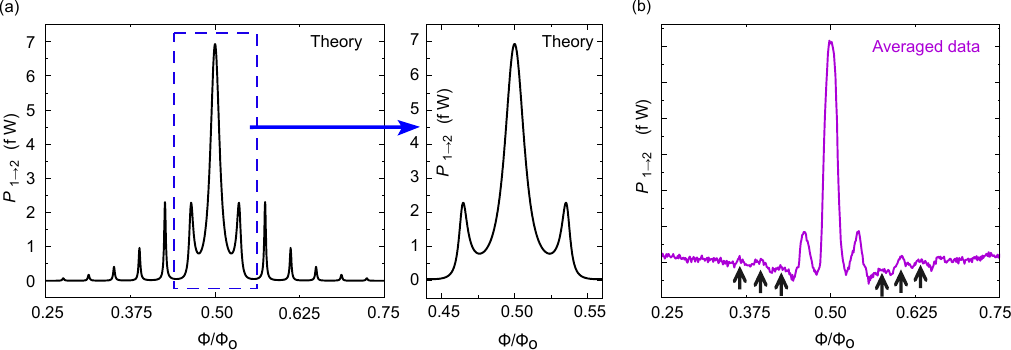}
	\caption{Observations in theory and experimental data. (a). The experimental results (left) and the theoretical prediction (right) for the power transmitted in the setup. The parameters in the calculation are as follows: $R= 6\,\Omega$, $L=0.8\,$nH, $I_p=30\,$nA, $f_{\rm q}^{(0)}=2\,$GHz, and $f_{\rm r}=7\,$GHz.
		\label{exp-theo-1}}
\end{figure}
\begin{eqnarray}\label{rhos}
	&&\rho_{0}=\frac{\Gamma_{2\rightarrow 0}\Gamma_{3\rightarrow 0}+\Gamma_{2\rightarrow 3}\Gamma_{3\rightarrow 0}+\Gamma_{2\rightarrow 0}\Gamma_{3\rightarrow 2}}{A+B+C}=\frac{A}{A+B+C}\nonumber\\
	&&\rho_{1}=0\nonumber\\
	&&\rho_{2}=\frac{\Gamma_{0\rightarrow 2}\Gamma_{3\rightarrow 0}+\Gamma_{0\rightarrow 2}\Gamma_{3\rightarrow 2}+\Gamma_{0\rightarrow 3}\Gamma_{3\rightarrow 2}}{A+B+C}=\frac{B}{A+B+C}\nonumber\\
	&&\rho_{3}=\frac{\Gamma_{0\rightarrow 3}\Gamma_{2\rightarrow 0}+\Gamma_{0\rightarrow 2}\Gamma_{2\rightarrow 3}+\Gamma_{0\rightarrow 3}\Gamma_{2\rightarrow 3}}{A+B+C}=\frac{C}{A+B+C}.
\end{eqnarray}
The power from Reservoir 1 to Reservoir 2 then reads
\begin{equation}\label{P-gen}
		P_{1\rightarrow 2}=\sum_{k,l} \rho_{k}\hbar\omega_{kl}\Gamma_{k\rightarrow l}^{(r=2)}.
\end{equation}
The calculated power $P_{1\rightarrow 2}(=-P_{2\rightarrow 1})$ is plotted in panel (a) of Fig. \ref{exp-theo-1} together with experimental results.

\vspace{1cm}

{\color{black}

\renewcommand{\thesubsection}{\arabic{subsection}}
\subsection{Flux qubit coupled to a bare resistor}

We can obtain the power nearly quantitatively at the symmetry point $\Phi=\Phi_0/2$, i.e. `` the central peak" by simply assuming that a bare resistor (reservoir) is coupled inductively to the qubit.
\begin{figure}[h!]
	\centering
	\includegraphics [width=0.35\textwidth] {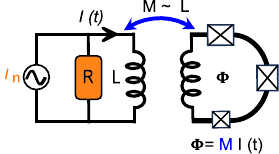}
	\caption{\textcolor{black}{Thermal noise of a resistor to a flux qubit via inductive coupling.}
		\label{fig1}}
\end{figure}
The equivalent circuit representing a resistor with resistance $R$ coupled via mutual inductance $M$ to a flux qubit is shown in Fig.~\ref{fig1}. Based on the Kirchhoff's circuit law, the current $I$ reads
\begin{equation}\label{I-1}
	I=\frac{R}{R+\imath \omega L}I_n
\end{equation}
The noise spectral density of \eqref{I-1} is then given by
\begin{equation}\label{SI-1}
	S_I(\omega)=\frac{R^2}{R^2+\omega^2 L^2}S_{I_n}(\omega).
\end{equation}
Since for the bare resistor with temperature $T=(k_B\beta)^{-1}$, $S_{I_n}(\omega)=\frac{2}{R}\frac{\hbar\omega}{1-e^{-\beta\hbar\omega}}$, we have
\begin{equation}\label{SI-2}
	S_I(\omega)=\frac{2R}{R^2+\omega^2 L^2}\frac{\hbar\omega}{1-e^{-\beta\hbar\omega}}.
\end{equation}
The flux into the qubit is $\Phi=M\,I$. Using Eq.\,\eqref{SI-2}, the noise sepectral density of flux, $S_\Phi=M^2\,S_I$, reads
\begin{equation}\label{SPhi-1}
	S_\Phi(\omega)=2M^2\frac{R}{R^2+\omega^2 L^2}\frac{\hbar\omega}{1-e^{-\beta\hbar\omega}}.
\end{equation}
Based on the golden rule, transition rate up, $\Gamma_\uparrow$, between the two levels of the qubit (ground state $|g\rangle$ and excited state $|e\rangle$) due
to the presence of the bath (resistor $R$) is given by
\begin{equation}\label{Gup-1}
\Gamma_\uparrow=\frac{1}{\hbar^2}|\langle g|\frac{\partial H}{\partial \Phi}|e\rangle|^2 S_\Phi(-\omega_{01}),
\end{equation}
where the Hamiltonian of the qubit is $\mathcal{\hat{H}}_{\rm q}=-\frac{\hbar}{2}\varepsilon \sigma_z$, which becomes equivalent to $\mathcal{\hat{H}}_{\rm q}=hf_{\rm q}(\Phi)\hat{b}^\dagger \hat{b}$ when restricted to the two lowest energy states. Here we have: $\varepsilon=\frac{2I_p}{\hbar}(\Phi-\Phi_0/2)$, with $I_p$ the persistent circulating current in the qubit loop. Since $\frac{\partial H}{\partial \Phi}=-I_p\sigma_z$, the transition rate is 
\begin{equation}\label{Gup-2}
	\Gamma_\uparrow=\frac{I_p^2M^2}{\hbar^2}\frac{2R}{R^2+\omega_{01}^2 L^2}\frac{\hbar\omega_{01}}{1-e^{-\beta\hbar\omega_{01}}}.
\end{equation}

In order to obtain the power between the two baths at temperatures $T_1$ and $T_2$, we need the diagonal elements of the density matrix. With no time-dependent driving we then have
\begin{equation}\label{densityM1}
	\dot{\rho}_{ee}=\Gamma_{\Sigma,\uparrow}\rho_{gg}-\Gamma_{\Sigma,\downarrow}\rho_{ee},
\end{equation}
where $\Gamma_{\Sigma,\uparrow}=\Gamma_{1,\uparrow}+\Gamma_{2,\uparrow}$ and $\Gamma_{\Sigma,\downarrow}=\Gamma_{1,\downarrow}+\Gamma_{2,\downarrow}$. By applying steady-state condition $\dot{\rho}_{ee}=0$, the population of the excited state reads
\begin{equation}\label{rhoee1}
	{\rho}_{ee}=\frac{\Gamma_{\Sigma,\uparrow}}{\Gamma_{\Sigma,\downarrow}+\Gamma_{\Sigma,\uparrow}}.
\end{equation}
The power, $\dot{Q}$, that goes out from the bath (resistor) is given by
\begin{equation}\label{power-1-2}
	\dot{Q}=\hbar\omega_{01}[\Gamma_{1,\uparrow}{\rho}_{gg}-\Gamma_{1,\downarrow}{\rho}_{ee}],
\end{equation}
where ${\rho}_{gg}=1-{\rho}_{ee}$ denotes the population of the ground state of the qubit. If we assume a two-bath system with $T_2\ll T_1$, then this leads to $\Gamma_{2,\uparrow}\simeq 0$, and $\Gamma_{2,\downarrow}\simeq \Gamma_{1,\downarrow}$. In this case power $\dot{Q}$ reads
\begin{eqnarray}\label{power-1-3}
	\dot{Q}&&\simeq \frac{1}{2}\hbar\omega_{01}\Gamma_{1,\uparrow},
\end{eqnarray}
We then have
\begin{equation}\label{power-1}
	\dot{Q}\simeq I_p^2M^2\frac{R}{R^2+\omega_{01}^2 L^2}\frac{\omega_{01}^2}{e^{\beta_1\hbar\omega_{01}}-1},
\end{equation}
where $\beta_1=(k_BT_1)^{-1}$ is the temperature of the hot bath (resistor). For the galvanic coupling here as we have here, $M\simeq L$. We then have
\begin{equation}\label{power-2}
	\dot{Q}\simeq I_p^2R\frac{(\omega_{01}L)^2}{R^2+(\omega_{01} L)^2}\frac{1}{e^{\beta_1\hbar\omega_{01}}-1},
\end{equation}
We see that for $L\ll R/\omega_{01}$, $\dot{Q}\propto L^2$ and it obtains its maximum value
\begin{equation}\label{power-max}
\dot{Q}_{\rm max}=I_p^2L\frac{\omega_{01}}{e^{\beta_1\hbar\omega_{01}}-1},
\end{equation}
when $L=R/\omega_{01}$. With the values used in the theoretical model, with the parameter values as stated above, we have from Eq.~\eqref{power-2}, $\dot{Q}\simeq 7.5~$fW at $T_{1} = 300$~mK, which remains in agreement with the experiment.
}

\vspace{0.2cm}

\clearpage
\twocolumngrid



\end{document}